\documentclass[aps,prl,twocolumn,preprintnumbers,amsmath,amssymb]{revtex4}

\usepackage{rotate}
\usepackage{epsfig}
\newcommand \be  {\begin{equation}}
\newcommand \bea {\begin{eqnarray} \nonumber }
\newcommand \ee  {\end{equation}}
\newcommand \eea {\end{eqnarray}}

\begin{document}


\title{Are defect models consistent with the entropy and specific heat
of glass-formers?}

\author{Giulio Biroli}

\affiliation{Service de Physique de Th{\'e}orique, Orme
des Merisiers, CEA Saclay, 91191 Gif sur Yvette Cedex, France;
}

\author{Jean-Philippe Bouchaud}
\affiliation{Service de Physique de l'{\'E}tat Condens{\'e}, Orme
des Merisiers, 
CEA Saclay, 91191 Gif sur Yvette Cedex, France;\\
and Science \& Finance, Capital Fund Management, 6-8 Bd Haussmann,
75009 Paris, France.
}%

\author{Gilles Tarjus}
\affiliation{Laboratoire de Physique Th{\'e}orique des Liquides, 4 Place Jussieu, 
75231 Paris Cedex 05, France
}%

\date{\today}
\begin{abstract}
We show that point-like defect model of glasses cannot explain 
thermodynamic properties of glass-formers, as for example the excess specific
heat close to the glass transition, contrary to the claim
of J.P. Garrahan, D. Chandler [Proc. Natl.  Acad. Sci. {\bf 100}, 9710
(2003)]. More general models and approaches in terms of extended defects are  
also discussed.
\end{abstract}
 
\maketitle
Kinetically constrained models ({\sc kcm}) display trivial 
thermodynamics but non trivial dynamics \cite{FA,KA,RS}. They encapsulate in 
a specific way the `free volume' idea according to which the slow dynamics of glasses
is due to rare, localized mobility defects \cite{freevolume,Glarum}. For most {\sc kcm}s, slow dynamics is indeed due
to defect motions \cite{RS}. A remarkable aspect of these models is that although the dynamical
rules are very simple and local, the emergent defect dynamics lead to
highly non trivial physical behaviour. As a consequence {\sc kcm}s
provide a theoretical framework for understanding many of the remarkable dynamical 
properties of fragile glass-formers, such as super-Arrhenius relaxation time \cite{RS,Evans,KAprl}, 
dynamical heterogeneities \cite{Harrowell,GC1,PNAS,FranzKA} and viscosity/diffusion decoupling \cite{decoupling}. 
That these models may in fact describe {\it quantitatively} the physical properties of glasses 
has been strongly advocated in a series of papers by
J.P. Garrahan and D. Chandler ({\sc gc}) \cite{GC1,PNAS}, and investigated 
further in \cite{Others}.

One of the major tenet of {\sc kcm}s is the complete decoupling
between dynamics and thermodynamics, at variance with the traditional
view of Adam and Gibbs \cite{AG}, and others \cite{KTW,BB}, in which the 
decrease of configurational
entropy with temperature is the fundamental underlying mechanism for the
viscous slowing down. It is therefore {\it a priori} surprising that 
{\sc kcm}s could have anything sensible to say about the entropy and 
specific heat of real materials -- even if these turn out to be acceptable
models of their dynamics. Nevertheless, the claim made in \cite{PNAS} is
that these models are also in quantitative agreement with the specific
heat of these materials. More precisely, assuming a perfect gas of free
defects, {\sc gc} give a formula for 
the specific heat at fixed pressure per particle of the liquid 
(in excess of that of the solid), which we reproduce from 
ref.\cite{PNAS} to be:
\be\label{cp}
\Delta C_p(T) = k_B\left(\frac{J}{T} \right)^2 c(T) {\cal N} + O(c(T)^2),
\ee  
where ${\cal N}$ is  the number of  physical molecules contributing to
enthalpy  fluctuations in a coarse-graining  cell, $J$ is the enthalpy
cost, in kelvins, for creating a mobile cell (remember  that liquids are considered
here  at constant  pressure,  hence the  importance  of distinguishing
between enthalpy and  energy). The concentration of mobile cells $c$ is {\it assumed} 
to be given by $c(T)/(1-c(T)) = A \exp(-J/T)$, where the numerical prefactor $A$ is 
argued by {\sc gc} to be rather large:  $\ln A$ is  akin  to  an   entropy gain  
$\Delta s_0/k_B$ associated with the creation  of a mobile cell.   
Note that there is a small  difference between the specific heat in excess to 
that of the  glass (considered in \cite{PNAS}) and that in excess of the crystal;
this difference is however irrelevant for  the present discussion.

In the following we first adress the results of \cite{PNAS} 
showing that even if the value  of  $\Delta C_p$ at  $T_g$  seems to  match experimental
values,   the  temperature  dependence    of   $\Delta  C_p$   predicted by
Eq. (\ref{cp}) is in total  disagreement with experimental values.
We then discuss  in  detail  the derivation of $\Delta C_p(T)$
and  the   physical assumptions  behind Eq. (\ref{cp}).  
This derivation leads to a formula different from Eq. (\ref{cp}) 
and makes clear that the defect contribution to $\Delta C_p(T)$
is in fact completely irrelevant close to the glass transition. In order 
to avoid a contradiction, one has to assume that the main contribution to
both the entropy and the specific heat comes from the {\it immobile} regions.
Although this is most reasonable physically, it implies that the models considered
by {\sc gc} are incapable of reproducing the thermodynamics of fragile glasses 
(and actually neither of strong glasses).
 
A clean way to test Eq. (\ref{cp}), that gets  rid of any ambiguity in
prefactors (we in fact disagree  with the  prefactor ${\cal  N}$ in
Eq. (\ref{cp}), see below), is to rewrite it as:
\be\label{cp2}
\frac{\Delta C_p(T)}{\Delta C_p(T_g)} \approx t^{-2}
\exp\left( \frac{J}{T_g} \frac{t-1}{t}\right)\qquad t=\frac{T}{T_g}. 
\ee 
This formula  is expected to be valid  for small $c \approx A \exp(-J/T)$,
where  the description of glassy   dynamics in terms of rare,  dilute,
defects could   make sense --  corresponding  to  a  large enthalpy of
creation  for mobility defects,  $J/T_g \gg  1$.

Using viscosity  data {\it at  fixed (atmospheric) pressure}, {\sc gc}
are able to estimate the value of $J/T_g$, found to be equal to $16.7$
for 3-bromopentane (3BP) and  $26.7$ for ortho-terphenyl (OTP), indeed
quite large compared to unity. Fig. 1, on the other hand, gives a plot
of ${\Delta  C_p(T)}/{\Delta C_p(T_g)}$,  obtained  from Eq.   (\ref{cp2}), as a
function   of $T/T_g$ for $J/T_g=10, 20,   30$ and compares  it to the
experimental  values  for 3BP \cite{3bp} and  OTP \cite{otp}.  (Note the
log  scale on  the $y$-axis.)  It  is clear  that  Eq.  (\ref{cp2}) is
totally incompatible with  the data when  $J/T_g$ is  large: it varies
much too  fast  with temperature,  and increases  with  $T$ instead of
decreasing (a similar criticism has been made by C. Angell and
coworkers \cite{Angell2} for excitation models). 

   An  acceptable     fit of   the    data  with
Eq. (\ref{cp2}) requires   $J/T_g \approx 1.3$,  meaning  that  within  this
description the density of defects is not small at all  -- but this of
course  is  then incompatible    with viscosity   data. For consistency,
Eq. (\ref{cp}) cannot be used for temperatures such that $c \simeq 1$; 
{\sc gc} restrict  their analysis to a  temperature range between $T_g$ and
$T_{1/2}$, where $T_{1/2}$ is defined  as the temperature at which the
logarithm  of the viscosity expressed  in Poise, or  of the relaxation
time  expressed  in  picoseconds  is  half its    value  at $T_g$ (see
Fig. 1). One can clearly see that, even in this restricted interval of
temperature close to $T_g$ (for which $c$ is of the order or less than
$10^{-1}$ \cite{PNAS}),  the  theoretical  prediction is in  complete
disagreement with the experimental data.
 

\begin{figure}
\begin{center}
\epsfig{file=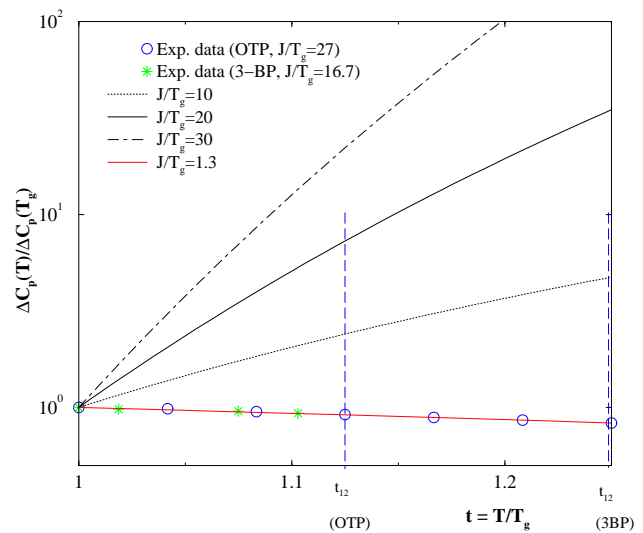,width=7cm,angle=270} 
\end{center}
\caption{Normalized excess specific heat as given by {\sc gc}'s formula,
 Eqs. (\ref{cp})   and (\ref{cp2}), for  $J/T_g=10, 20,  30$, compared
 with the   experimental data for OTP   and 3-BP, for   which {\sc gc}
 estimate,     from  viscosity    data,   $J/T_g=26.7$  and     $16.7$, 
 respectively. Note the log  scale on the  $y$-axis. The two  vertical
 lines correspond to $T_{1/2}$ defined as the temperature at which the
 logarithm of the viscosity  expressed in Poise,  or of the relaxation
 time expressed in picoseconds is half its value at $T_g$. }
\label{Fig1}
\end{figure}

The previous discussion suggests that the thermodynamic predictions 
of the models considered in \cite{PNAS} are incompatible with experiments. 
As we discuss below, there is in fact no real contradiction:
the order of magnitude of the defect contribution to the specific heat 
is simply much smaller than 
the experimental value between $T_{g}$ and $T_{1/2}$,
which is instead dominated by the thermodynamic properties of immobile regions. 
In the models considered in \cite{PNAS,GC1}, however, the immobile regions give 
a trivial contribution to thermodynamics properties. In order to be
consistent, one should assume that these regions dominate the 
thermodynamics but are irrelevant dynamically.

Let us now  reconsider in detail the derivation  of Eq. (\ref{cp})  taking
explicitely  into account that  experimentally {\it both} the specific
heat and the  excess  entropy  are found to  be of order $k_B$   per
particle or  larger   at the glass   temperature -- indeed a  very important
``stylized fact'' of fragile glasses. An immediate consequence of
this experimental fact is that the excess entropy of the liquid cannot
be due to point-like defects only: since these must be dilute to lead
to  large relaxation times, their  contribution  to the excess entropy
must be small. Within a  simple coarse-grained defect description, the total
excess entropy per cell (with respect to the crystal), $\Delta s$, is:
\be\label{s}
\Delta s = -k_B [c \ln c + (1-c) \ln (1-c)] + c \Delta s_0 + \Delta s_c, 
\ee
where the  first term corresponds to the  entropy of the ideal  gas of
defects, the second   term to the entropy difference  between
mobile and  immobile cells (see above), and  the   last term  is  the
configurational entropy associated with frozen cells \cite{footnote}.
In order to get  the experimental order  of magnitude of $\Delta  s$ at
$T_g$ and at  the  same  time  have  $\Delta  C_{p}$  dominated by   defect
properties,  one must assume that (i) $\Delta s_c$   is of order ${\cal N}
k_{B}$, hence  much larger than the  defect entropy itself; (ii) $\Delta
s_c$ must then be  temperature  independent, otherwise it would  give  the
leading contribution  to  $\Delta C_{p}$. Although large, the
excess entropy $\Delta s_c$ should not play any  physical role in the slowing
down of the material in this picture (in strong contrast with the view expressed in \cite{AG,KTW,BB}). 
Therefore, the excess specific heat at constant pressure 
{\it per cell}, $\Delta c_p(T) = T {d\Delta s(T)}/{dT}$, is given by:
\be
\Delta c_p(T)  = k_B \frac{J}{T}  T \frac{dc(T)}{dT}  \approx k_B(J/T)^2 c(T) ,
\ee
where  we have used $c(T) \approx \exp(\Delta s_0-J/T)$,  valid in the limit $c \ll
1$. 
Note that the functional form of $c (T)$ cannot be derived
within the model and it is in principle obtained from the coarse-grained procedure
that should map glass-forming liquids to {\sc kcm}s.
Since until now no such procedure has been developed one 
has to postulate an a priori functional form $c (T)$.
The simple physical assumptions described above correspond to the one assumed in \cite{PNAS}. 
However, we will show below that our conclusion is indeed independent of the form of $c (T)$.
We have obtained a specific heat {\it per cell} given by $k_B(J/T)^2
c$, and therefore a specific heat {\it per molecule} given by:
\be\label{cp3}
\Delta C_p(T) \approx k_B \left( \frac{J}{T}\right)^2 c(T) {\cal N}^{-1}, 
\ee  
a factor ${\cal N}^2$ smaller than Eq. (\ref{cp}). The same expression
is obtained by looking directly at the  enthalpy fluctuations \cite{footnote2}.  Taking
very favorable values for the parameters $c =  0.01$, $J/T_g = 30$ and
${\cal N}=10$ leads to  $\Delta C_p(T_g) \approx 0.1  k_B$, already a  factor 100
too small compared to experimental values ($9 k_B$  for 3-BP, $13 k_B$
for OTP).  The only way to avoid a contradiction is that the temperature dependence of $\Delta  s_{c}$
is strong, and that in fact all the thermodynamics is contained in that very contribution.

The above conclusion is based on comparing the experimental specific heat to that predicted 
by defects only. A more general argument (see also \cite{BB,LW1}), independent of the shape of $c(T)$, relies on  
the entropy change between -- say -- $T_g$ and $T_{1/2}$, which is experimentally found to
found to be of order $k_{B}$ ($\approx ~4k_{B}$ for OTP). 
Thus, not only one finds that $\Delta s_c(T)$ has to be much larger than 
$(-k_B [c \ln c + (1-c) \ln (1-c)]+c \Delta s_0)$
at $T_{g}$, as discussed previously, but also that $\Delta
s_c(T)$ has to provide the leading contribution between $T_{g}$
and $T_{1/2}$ since in this regime the ideal gas entropy of defects \cite{footnote3} is
negligible \cite{PNAS}. As a consequence, using the relationship 
$\Delta c_p(T) = T {d\Delta s(T)}/{dT}$, one finds again that
the main contribution to $\Delta C_{p}$ comes from the {\it immobile regions} close to $T_{g}$. 
Note that this result is independent of the functional form $c (T)$, 
as long as defects are dilute, $c \ll 1$, a necessary condition 
for the {\sc gc} description to be useful \cite{footnoteT}.

So, are all {\sc kcm}s (or more generally defect models) doomed to fail in describing thermodynamics? 
The point of view advocated in \cite{GC1,PNAS} is that after having
appropriately coarse-grained time and space a liquid can be strictly 
speaking considered as a {\sc kcm} with trivial thermodynamics and
simple kinetic constraints that induce slow dynamics. As we have shown, this 
line of thought is clearly insufficient to recover thermodynamics. 
Another point of view (see for example \cite{KA,KAprl}) 
is to only assume that the effective dynamics for a lattice model of a glass-forming liquids is
characterized by some kinetic constraints. In this case
the trivial thermodynamics is not necessary but just a choice of simplicity. A quantitative
model (arising from a real mapping from liquid dynamics) would also contain 
effective interactions between particles and would therefore lead to non-trivial thermodynamics.
One important fact remains though: even if the thermodynamics can be
non trivial, it is only an indirect cause of slow dynamics, through the 
temperature dependence of the defect density, and not a driving cause of the slowing
down. 
In our view, any eligible theory
of the  glassy state has to  produce a convincing explanation  of the
remarkable connection between thermodynamics and  dynamics, more
precisely between  the configurational entropy and the relaxation
time \cite{Angell}. At the moment this seems to be an important missing
piece in {\sc kcm}s. Hopefully, future works on {\sc kcm}s and their possible
generalisations will unveil whether or not some of these models can pass this
important test. A way to increase the entropy contribution of
mobility defects and find a relationship  between dynamics  and
thermodynamics seems to be through {\it extended defects}. This is
the path followed (using different arguments) in \cite{AG,SN,KTW,KT,BB,W2} 
(see also \cite{CMPP,Sch} for quantitative
computation of the excess configurational entropy in model systems). 
In all  these scenarii, the presence of mobility  regions,  that are
typically interfaces (e.g. domain walls), is
driven by  a thermodynamic  mechanism  with a (possibly avoided) critical point.

\begin{acknowledgments}
We thank C. Alba-Simionesco, L. Berthier and D.R. Reichman for interesting discussions.
GB is partially
supported by the European Community's Human Potential Programme
contracts HPRN-CT-2002-00307 (DYGLAGEMEM).
\end{acknowledgments}

\vskip0.4cm
{\bf \large Addendum}\\
In the following reply, Chandler and Garrahan have amended their original
model which now allows for two different species of excited cells -- most
of them are in this new version immobile, and very few are mobile.
Even though the microscopic justification of this ad hoc assumption is
unclear, the new fitting parameter, the ratio of excited mobile to excited
immobile cells, indeed allows one to obtain both a large specific heat
jump and a large relaxation time. In such a framework, however,
the strong connection between thermodynamical and dynamical fragility,
which we emphasized above, becomes completely out of reach or very
artificial (thermodynamics is dominated by immobile excited
cells that are irrelevant as far as dynamics is concerned). We also note
that despite the introduction of another fitting parameter
[the temperature dependence of the ratio of excited mobile to excited
immobile cells] the theory proposed by Chandler and Garrahan still  leads
to a specific heat which has the wrong temperature dependence  -- note
that the scale of the y-axis in their Fig. 1 makes it hard  to see that
their prediction for the excess of specific heat is in fact increasing
with temperature, in contradiction with experiments that
clearly show an opposite trend.
It might be possible, by introducing additional assumptions and fitting
parameters, to obtain yet another version of Chandler and Garrahan's
model that "accounts for" these basic experimental facts. However, all this
is at the expense of the simplicity, the predictive power and, more
importantly, the very meaning of the underlying physical picture.


\begin{thebibliography}{99}

\bibitem{FA}
G.H. Fredrickson and H.C. Andersen, Phys. Rev. Lett. {\bf 53}, 1244 (1984);
J. Chem. Phys. {\bf 83},5822 (1985). G.H. Fredrickson and S.A. Brawer
J. Chem. Phys. {\bf 84}, 3351 (1986).

\bibitem{KA}
W. Kob and H.C. Andersen, Phys. Rev. E {\bf 48} (1993) 4364.

\bibitem{RS} for a review, see F. Ritort, P. Sollich, Adv. Phys. {\bf 52}, 219 (2003).

\bibitem{freevolume} M.H. Cohen and G.S. Grest, Phys. Rev B {\bf  20},
1077 (1979); {\bf  21} 4113 (1980).

\bibitem{Glarum} S.H. Glarum, J. Chem. Phys. {\bf  33}, 639 (1960) 

\bibitem{KAprl}
C. Toninelli, G. Biroli, and D. S. Fisher
Phys. Rev. Lett. {\bf 92}, 185504 (2004) and cond-mat/0410647.

\bibitem{Evans} P. Sollich , M.R. Evans, Phys. Rev. Lett {\bf 83}, 3238 (1999).

\bibitem{Harrowell} P. Harrowell
Phys. Rev. E {\bf 48}, 4359 (1993).
M.M. Hurley, P. Harrowell, Phys. Rev. E {\bf 52}
1694 (1995) and refs therein.

\bibitem{GC1} J.P. Garrahan, D. Chandler, Phys. Rev. Lett. {\bf 89}, 035704 (2002). 
	
\bibitem{PNAS} J.P. Garrahan, D. Chandler, Proc. Natl. Acad. Sci. {\bf 100}, 9710 (2003). 


\bibitem{FranzKA}
S. Franz, R. Mulet and G. Parisi, Phys. Rev. E {\bf 65}, (2002) 021506.
 

\bibitem{decoupling}  Y. Jung, J.P. Garrahan, D. Chandler,
Phys. Rev. E, {\bf  69}, 061205 (2004). {\it Length scale for the
onset of Fickian diffusion in supercooled liquids},  L. Berthier, 
D. Chandler, J. P. Garrahan, cond-mat/0409428.

\bibitem{Others} L. Berthier, J.P. Garrahan, Phys. Rev. {\bf E 68}, 041201
 (2003);     S.    Whitelam,     L.     Berthier,   J.P.     Garrahan,
 Phys. Rev. Lett. {\bf 92}, 185705 (2004); L. Berthier, J.P. Garrahan,
 {\it Numerical  study  of   a fragile three-dimensional   kinetically
 constrained model}, cond-mat/0410076.

\bibitem{AG} G. Adam, J. H. Gibbs, J. Chem. Phys. {\bf 43} 139 (1958).

 \bibitem{KTW} T. Kirkpatrick, P. Wolynes, Phys. Rev. B {\bf 36}, 8552
 (1987);  T.  R.  Kirkpatrick, D.  Thirumalai,   P.  G. Wolynes,  {\it
 Phys. Rev.  A}   {\bf 40}  (1989) 1045;   X.   Xia,  P.  G.  Wolynes,
 Proc. Nat.  Acad. Sci. {\bf  97},  2990 (2000), Phys.  Rev. Lett {\bf
 86}, 5526 (2001).
	
\bibitem{BB} J. P. Bouchaud, G. Biroli, J. Chem. Phys. 121, 7347 (2004)


\bibitem{3bp} S. Takahara, O. Yamamuro, T. Matsuo, J. Phys. Chem. {\bf
 99} (1995) 9589.

\bibitem{otp} S.S. Chang and A.B. Bestbul, J. Chem. Phys. {\bf  56} (1972) 503. 

\bibitem{Angell2}  R. Moynihan, C. A. Angell, J. Non. Cryst. Solids {\bf 274}
, 131   (2000); C. A.   Angell, J. Phys.  Cond. Matter  {\bf 12}, 6463
(2000).

\bibitem{LW1} V. Lubchenko and P.G. Wolynes, J. Chem. Phys. {\bf 121}, 2852 (2004).

\bibitem{footnote} Even if
the difference between  the   heat capacities  of  the glass  and  the
crystal is small,  the entropy of the  glass, considered as a ``frozen
liquid'', is  significantly higher than that  of the crystal down to 0
K.

\bibitem{footnote2} A very direct way to obtain
Eq. (\ref{cp3}) is writing the enthalpy per particle due to mobile 
cells $H=Jc (T)/{\cal  N}$ and using the thermodynamic relation 
$\Delta C_{p} (T)=dH/dT$.

\bibitem{footnote3} Even if  there was  a static interaction
between defects, the entropy contribution cannot be larger than  the
ideal gas  one. 

\bibitem{footnoteT} This in principle does not directly apply to 
strongly cooperative defect models. Their structural relaxation
timescale (and inverse of the defect density) can increase extremely fast
when the temperature decreases or the particle density increases \cite{RS,KA,KAprl}.
But at the same time their entropy decreases mildly and smoothly
in such a way that it is not necessary to have
a very small entropy in order to have a very large relaxation time
\cite{Sellitto}.

\bibitem{Sellitto} M. Sellitto, J. Phys. Condens. Matter {\bf  12} (2000) 6477.

\bibitem{Angell} C. A. Angell, Jour. Res. NIST, {\bf 102} 171 (1997); 
 R. Richert and C. A. Angell, J. Chem. Phys. {\bf 108} 9016 (1998).

\bibitem{SN}
S. Sachdev and  D. R. Nelson, Phys.  Rev. Lett. {\bf 53}, 1947 (1984);
Phys. Rev. B {\bf  32},  1480 (1985).  D. R.   Nelson and F.  Spaepen,
Solid State Phys. {\bf 42}, 1 (1989).
 
 \bibitem{KT}  D. Kivelson, S. A. Kivelson,  X.-L.  Zhao, Z. Nussinov,
 and G.   Tarjus,  Physica A {\bf   219},  27  (1995); G.   Tarjus and
 D. Kivelson, J.  Chem. Phys. {\bf 103},  3071 (1995); ibid 
J.  Chem. Phys. {\bf 109},  5481 (1998); D. Kivelson and
 G. Tarjus, Phil. Mag. B{\bf 77}, 245 (1998);  

\bibitem{W2} V. Lubchenko and P.G. Wolynes,
J. Chem. Phys. {\bf  119} 9088 (2003), J. Chem. Phys. {\bf  121}, 2852 (2004),
Phys. Rev. Lett. {\bf  87} , 195901 (2001);
X. Xia and P.G. Wolynes PNAS {\bf  97}  2990 (2000) and refs. therein.


\bibitem{CMPP} B. Coluzzi, M. M{\'e}zard, G. Parisi, P. Verrocchio,
J.Chem.Phys. {\bf  111}, 9039 (1999).

\bibitem{Sch}  K.-K. Loh, K. Kawasaki, A. R. Bishop, T.
	Lookman, A. Saxena, J. Schmalian, Z. Nussinov,
	{\it Glassy behavior in systems with Kac-type step-function
	interaction}, cond-mat/0206494

\end{thebibliography}
\end{document}